\documentclass{aa}
\usepackage{graphics}
\usepackage{epsfig}

\begin{document}

\thesaurus{02.01.2 - 08.14.2 - 02.09.1 - 02.13.1 - 08.09.3}

\title{Repetitive rebrightening of EG Cancri: evidence for viscosity 
decay in the quiescent disk?}

\author{
Yoji Osaki\inst{1}, Friedrich Meyer\inst{2}, and Emmi Meyer-Hofmeister
\inst{2}}

\offprints{Yoji Osaki; osaki@net.nagasaki-u.ac.jp}

\institute{Faculty of Education, Nagasaki University, Nagasaki 852-8521, Japan 
\and
 Max-Planck-Institut f\"ur Astrophysik, Karl-Schwarzschild Str. 1, \\
D-85740 Garching, Germany}

\date{Received; accepted}

\authorrunning{Yoji Osaki et al.}
\titlerunning{Repetitive rebrightening of EG Cancri}

\maketitle

\begin{abstract} 

A WZ Sge-type dwarf nova, EG Cancri, exhibited six consecutive 
mini-outbursts with a mean interval of about seven days after the end of 
the main outburst in 1996/1997. Most unusual was that the star 
abruptly entered into a deep faint minimum after such frantic 
activities. We propose that this peculiar phenomenon may be understood 
in terms of viscosity decay in the cold disk. In this picture, the
viscosity is produced  by MHD turbulence due to the magneto-rotational
instability ('Balbus-Hawley instability') and dies down
exponentially with time when the disk becomes cold because the
magnetic fields decay due to finite conductivity in the cold disk
(Gammie \& Menou 1998). But the viscosity is refreshed to a high value
every time when a mini-outburst occurs (i.e., the disk becomes hot
again). It is argued that a sudden cessation of repetitive
mini-outbursts may be brought about by a very small reduction in
viscosity or a small increase in its decay rate, which may in turn be
produced most likely by stochastic fluctuations of magnetic fields.
Numerical simulations based on a simple model reproduce the observed
light curve of EG Cancri very well. We discuss possible causes why the 
reflares after the main outburst occur mostly in the WZ Sge-type stars. 


\end{abstract}

\section {Introduction}  

  WZ Sge stars are a small group of dwarf novae, exhibiting 
large-amplitude outbursts (6-8 mag instead of 2-5 mag) and very long 
recurrence times (decades instead of months), compared with typical 
dwarf novae (Bailey 1979, Downes 1990, O'Donoghue et
al. 1991).  They 
also belong to the SU UMa-type sub-class of dwarf novae, which exhibit the 
two different types of outbursts, a short 'normal outburst' (with a 
typical outburst duration of a few days) and a long 'superoutburst' 
lasting for more than ten days. The superoutbursts are almost always 
accompanied by 'superhumps'. An appearance of superhumps 
during an outburst is taken as evidence for a SU UMa-type dwarf 
nova. These superhumps are periodic humps in the photometric light
curve with periods slightly longer than the orbital period of the binary. 

 EG Cancri, a WZ Sge-type dwarf nova, underwent a large outburst in 
December 1996 after 19 years of dormancy (Matsumoto et al. 1998b,  
Patterson et al. 1998). The WZ Sge-type stars often show rebrightening 
or reflares after the end of the main outburst (Richter 1992),
e.g., in the 1978 outburst of WZ Sge (Patterson et al. 1981),
the 1961 outburst of AL Com and the 1995 outburst of the same star ( 
Kato et al. 1996,
Howell et al. 1996, Patterson et al. 1996). Thus
EG Cnc was closely watched by many observers. EG Cnc
in fact made reflares after the end of the main outburst.  However, 
what was unexpected is that EG Cnc made reflares not only once but  
exhibited six consecutive mini-outbursts with a mean interval of 7
days. The most unusual feature was that the star abruptly stopped the
outbursts after the sixth mini-outburst and entered into a deep faint
minimum. The light curve of EG Cnc can be found from
VSNET (http:/www.kusastro.kyoto-u.ac.jp) or in the papers of Kato et
al. (1997), Matsumoto et al. (1998a) and Patterson
et al. (1998). Fig. 1 shows the eruption light curve as obtained
by Patterson et al. (1998) from all available data.

\begin{figure*}[ht]
\includegraphics[width=16.cm]{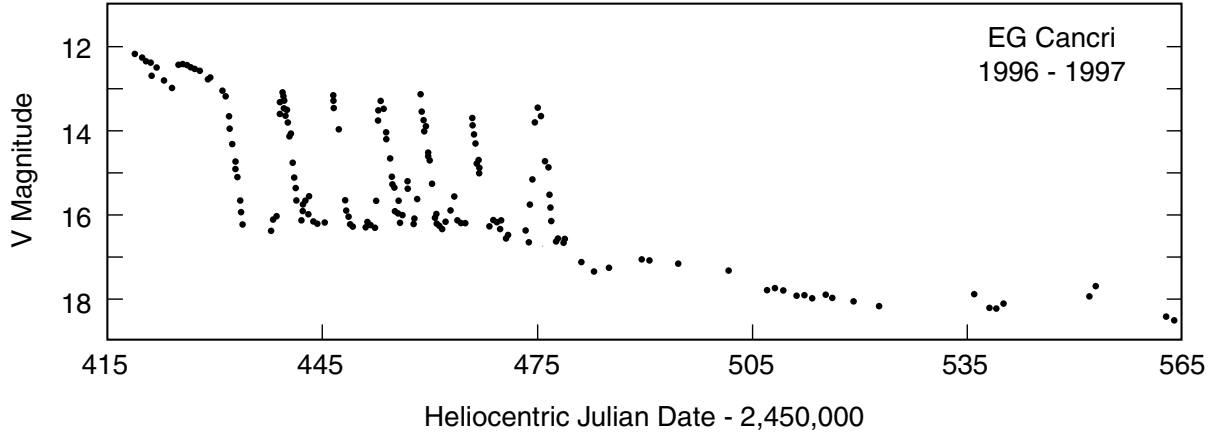}
\caption{Outburst light curve of EG Cnc from Patterson et al. (1998).}
\end{figure*}

  As for theory dwarf nova outbursts are now well understood in terms
of the disk instability model (see reviews of Cannizzo 1993 and
Osaki 1996) in which the accretion disk undergoes a 
thermal relaxation oscillation between a hot ionized state (corresponding 
to outburst) and a cold unionized state (corresponding to quiescence). 
Based on the disk instability model, outburst light curves of dwarf 
novae have been simulated by various workers using an $\alpha$  
parametrization (Shakura \& Sunyaev 1973) for the 'turbulent'
viscosity. It has, 
however, long been recognized (Smak 1984) that we must choose 
different values for viscosity $\alpha_{\rm hot}$ in the hot state
and $\alpha_{\rm cold}$ in cold state, i.e., something like $\alpha_{\rm 
hot} \sim 0.2$ and  $\alpha_{\rm cold} \sim 0.02$. It has, however, been 
pointed out by various investigators (Smak 1993,
Osaki 1994, 1996, Howell et al. 1995, Meyer
et al. 1998) that in the case of  WZ Sge 
stars an extremely low viscosity in quiescence, as low as $\alpha_{\rm 
cold} \sim 0.001$ is required to explain their long recurrence times.

  The magneto-hydrodynamic origin of the turbulent viscosity in the
accretion disk has long been suspected. This idea was put on firmer ground with
analysis and recent results of  computations
for the magneto-rotational instability (Balbus and Hawley 1991,
Brandenburg et al. 1995, Hawley et al. 1996,
Armitage 1998). Originally this
instability was discovered by Velikov (1959) and Chandrasekhar
(1961).
Gammie \& Menou (1998) suggested that in the quiescence the MHD
turbulence could decay because of the poor conductivity of the cold disk. 
Meyer \& Meyer-Hofmeister (1999) suggested the following picture for
the viscosities in the hot and cold state. The magneto-rotational
instabilty of small-scale magnetic fields in the hot disk produce the
high viscosity. In quiescence this is not possible anymore (Gammie \&
Menou 1998), but the magnetic fields reaching over from
the companion star allow a weaker MHD turbulence and cause the
lower viscosity in the cold disk. The extreme low viscosity in WZ Sge
stars then may be due to the absence of magnetic fields from the
companion star in the case of cool degenerate secondary stars.

  For the explanation of the repetitive rebrightening of EG Cnc, a few 
possibilities were suggested. The first is that the observed six
mini-outbursts were definitely of the 'normal outburst' variety as  
indicated by the observational evidence (i.e., duration of outburst,
spectroscopic evidence, and no photometric evidence of superhumps).
Osaki et al. (1997) suggested that mini-outbursts could be produced
by a temporal enhancement of viscosity in the quiescent disk just
after the main outburst. They demonstrated that the mini-outbursts
and their sudden cessation observed in EG Cnc could be reproduced
if the viscosity in the quiescent disk was kept high just after the
main outburst for the time of the mini-outbursts but then 
suddenly dropped to the extremely low value of the pre-outburst stage. 
However, no physical explanation was offered why the viscosity in the
quiescent disk of EG Cnc varied in such a way. 

Another possibility might be a temporal enhancement of the mass transfer 
rate from the secondary star due to irradiation of the 
secondary's atmosphere by the heated white dwarf primary star. This 
possibility might arise from the fact that the six mini-outbursts look
like the rapid outbursts observed for ER UMa stars, a subclass of SU
UMa stars which are thought to have mass transfer from the
secondary stars on high rate (see also Osaki 1996, Kato et al. 1999).
In fact, Hameury et al. (2000) made simulations of dwarf novae testing
the effect of secondary illumination and disk illumination with
several different parameters and commented that one of the simulated
light curves looked like the observed rebrightening of EG Cnc.
However, their light curve looks like that of damped oscillations
which is quite different from the rebrightenings in EG Cnc.
Furthermore, if the secondary star were illuminated, a periodic 
signal corresponding to the binary orbital period would have appeared. 
But no such evidence was known from observations. The disk illumination 
is also unlikely because the quiescent disk during six mini-outbursts 
was rather cool as evident from its red color (Patterson et al. 1998).

 The most unusual feature of the outburst of EG Cnc was that the star
executed six mini-outbursts with a fairly constant level of 
maximum luminosity with more or less similar outburst intervals, but 
then abruptly went down into a deep faint minimum. In this paper, we
address this particular feature of the light curve of EG Cnc by 
proposing a possible model in Sect.2. We then present numerical 
simulations based on our simplifying model in Sect.3. Finally discussions 
are given in Sect.4 and conclusions in Sect.5 .

\section {A working model}
\subsection {The standard disk instability model}

In our model the six mini-outbursts of EG Cnc are taken to be 
normal outbursts produced by the thermal disk instability. The disk 
instability model explains the dwarf nova outbursts as  
thermal relaxation oscillations between a hot fully ionized state and
a cold unionized state in the bi-stable disk. The thermal relaxation
oscillation is most easily understood locally in terms of the
so-called S-shaped equilibrium curve. Fig 2a shows the hot upper and
cool lower branch connected by the partially ionized unstable branch.
At each radius in the disk there exist two critical surface densities,
$\Sigma_{\rm max}$, above which no cold equilibrium state exists and
$\Sigma_{\rm min}$, below which no hot state exists. At the end of an
outburst, all parts of the disk return back to a cold state and the
surface density lies between $\Sigma_{\rm max}$ and $\Sigma_{\rm min}$
as illustrated schematically in Fig. 2b. Matter from the secondary
and viscous evolution within the disk makes the surface density to
increase in the disk. The surface density will ultimately reach 
the local critical density $\Sigma_{\rm max}$ at some point in the disk. 
A heating transition to the hot state results (see Fig. 2c). The 
heating front propagates both inward and outward, turning all of the 
disk matter into the hot state. During the hot state mass flows inward
at a high rate and the surface density decreases. When
$\Sigma_{\rm min}$ is reached the disk turns to the cool state. The
cooling transition always occurs at the outer edge of the disk and
the cooling front propagates inwards, extinguishing the outburst. 

\begin{figure*}[ht]
\includegraphics[width=16.cm]{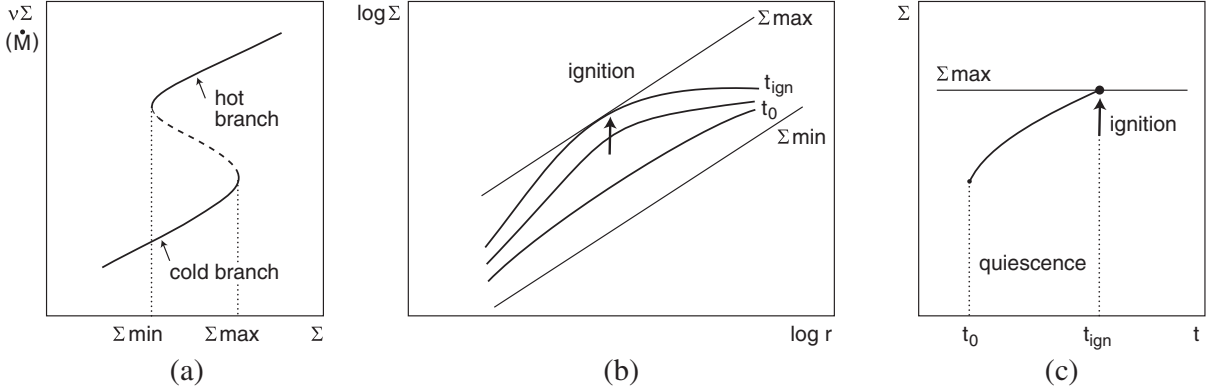}
\caption {(a) A schematic S-shaped thermal equilibrium curve showing the
two critical surface densities, $\Sigma_{\rm min}$ and $\Sigma_{\rm max}
$. (b) Time evolution of the surface density in the quiescent disk. (c) 
Time evolution of surface density at the ignition point.}
\end{figure*}

It is well known that for the outburst two different possibilities
exist as to where the disk first reaches the ignition point, either at
the outer edge or in the inner part of the disk. Accordingly
two types of outbursts are known, the outside-in outburst and the inside-out 
outburst. An outside-in outburst results if the mass transfer rate
from the secondary star is high while the inside-out 
outburst occurs if the viscous evolution within the disk is effective.     

 Rapid repetition of mini-outbursts with a recurrence time as short as 
seven days as observed after the end of the main outburst in EG Cnc 
can be understood due to either an increased mass transfer rate from the 
secondary star or to an enhanced viscosity in the disk. The
model of Osaki et al. (1997) for EG Cnc was based on the
second possibility.

Let us now describe our working model for the whole outburst cycle of WZ 
Sge stars, in general, and, EG Cnc, in particular, from the stand-point 
of temporal evolution of the disk viscosity. We assume that the viscosity 
in the hot state, $\alpha_{\rm hot} \sim 0.2$, is produced by dynamo action
by the MHD turbulence due to the 'Balbus-Hawley instability', but in the cold
quiescent disk these magnetic fields disappear due to poor
conductivity as described in Sect. 1. The very low viscosity in
quiescence in WZ Sge stars, $\alpha_{\rm cold} \sim 0.001$, may then
be produced by some other mechanisms such as tidal spiral density
waves (see, e.g. Spruit 1989) or some other hydrodynamic
instabilities ( e.g. Papaloizou \& Pringle 1984). With such low
viscosity the mass transferred from the secondary will be stored
simply in a torus at the 'Lubow and Shu radius' (Lubow \& Shu 1975),
the circularization radius, and the mass accumulated will be very
large when the next outburst is ignited. The outburst then is
characterized by a large amplitude and long duration due to
the large amount of mass accumulated during the long quiescence
(see, e.g., Osaki 1995a). 

\subsection {The effect of decreasing viscosity}

At the end of the outburst the viscosity will only gradually
decrease due to the finite resistive decay time of the magnetic fields
and the viscosity parameter may still remain high, as large
as $\alpha_{\rm cold} \sim 0.1$ in the cold state in the beginning quiescence. 

Let us now consider what happens when the viscosity starts  to
decrease. If the viscosity in the cold disk would stay 
high all the time,  the surface density will sooner or later reach the 
local critical density $\Sigma_{\rm max}$ at some point in the disk.
and a normal outburst will be ignited as illustrated in Fig. 2. 
However if the viscosity decreases exponentially with 
time due to magnetic diffusivity, the critical surface density 
$\Sigma_{\rm max}$ increases exponentially  since $\Sigma_{\rm max}$
is proportional to about $\alpha_{\rm cold}^{-0.8}$. In this 
case a competition occurs between the local increase in surface density and 
the increase of the critical surface density, illustrated in 
Fig. 3. This situation is comparable to that of a runner who tries to reach 
the goal, but the goal is receding. If the runner wins, we have the 
ignition to normal outburst (Fig.3a). If the runner once loses, he
never makes 
it because the goal recedes increasingly faster. This is illustrated
in Fig. 3b as the surface density curve is convex while the critical 
surface density curve is concave.

\begin{figure}[ht]
\includegraphics[width=5.cm]{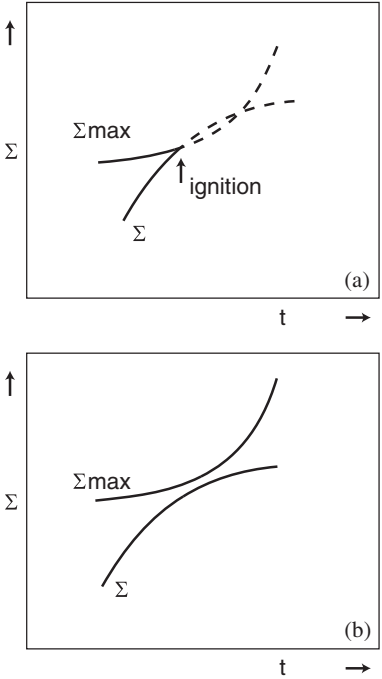}
\caption {Schematic diagrams for evolution of the surface density $\Sigma$
with time together with the critical surface density
$\Sigma_{\rm max}$, which increases if a viscosity decay is taken into
account.
(a) situation in a case where ignition occurs, (b) when the critical
surface density is not reached and no ignition occurs. }
\end{figure}

We further propose that the viscosity returns back to the original high 
value every time a mini-outburst occurs since the disk 
becomes hot again and fully ionized and the high 
electric conductivity is restored. The 'Balbus-Hawley instability' becomes 
effective again, reproducing MHD turbulence and high viscosity. After
the end of a mini-outburst, almost the same situation will be realized
in the disk as before and a new competition begins. An additional
fact influences the outcome if this competition:  the dynamo
generation  of magnetic fields by the 'Balbus-Hawley instability'
is stochastic and in such a case the  strength of regenerated magnetic
fields and their configuration may not be the same after every
outburst. The natural consequence of this is that there are 
some fluctuations in the resultant viscosity and its decay time
fluctuates from mini-outburst to mini-outburst.
If now the competition between the viscous diffusion of disk 
matter and the viscosity decay is very delicate, one time the diffusion 
of matter wins and an outburst occurs but another time the decay of 
viscosity wins. If then the increase of surface density fails to reach the 
critical one, it never catches up again since the viscosity decay is a 
run-away process and the disk viscosity becomes very low and the star 
enters into a long phase of faint minimum. We postulate 
that the six such mini-outbursts are repeated in this way in EG Cnc before 
a final competition leads to a faint deep minimum.        

We now estimate the characteristic decay time of magnetic fields and 
of the viscosity due to the finite conductivity in the cold disk matter. 
Hawley et al. (1996) found in numerical simulations that the dynamo
created magnetic fields are no longer sustained and start to decay 
when the magnetic Reynolds number $Re_{\rm m}$ becomes less than
$\approx 10^4$. Here the magnetic Reynolds number is $Re_{\rm m}=Lv/\eta$  
where $L$, $v$, and $\eta$ are the characteristic length scale, the 
characteristic velocity, and the magnetic diffusivity, respectively.  We 
adopt $L \sim H$ and $v \sim c_s$ in the accretion disk, where $H$  
is the vertical scale height of the disk and  $c_s$ the isothermal sound 
velocity. For the decay time $\tau_{\rm decay}$ of 
magnetic fields we then obtain 
$$\tau_{\rm decay} \sim L^2/\eta \sim Re_{\rm m} \cdot H/c_s 
\sim Re_{\rm m} \cdot {1 \over \Omega}, $$
where $\Omega$ is the Keplerian rotation frequency of the disk matter. 
The magnetic Reynolds number in the cold disk is estimated to be 
$Re_{\rm m}\approx 10^3\, - \,10^4$ (Gammie \& Menou 1998).
In the disks of WZ Sge stars the inverse of the Keplerian angular frequency  
is of the order of 300 sec. This yields the characteristic decay time
of order of a week to a month. A time scale of this order 
seems to be right for our model, as the repetition time of 
mini-outbursts is about a week.

On the other hand, once the disk becomes hot and ionized, the 
'Balbus-Hawley instability' becomes effective and starts to build up the 
magnetic fields. Their growth time is thought to be very short,
estimated to be of the order of the inverse of 
the Keplerian rotation frequency, $\tau_{\rm growth} \sim 
{1/\Omega}$.  This time-scale is sufficiently short that the decayed 
magnetic fields can be brought back to the original level and that thus the 
viscosity is refreshed to the original high value every time when a 
mini-outburst occurs. 

This large disparity of growth rate and decay rate of magnetic fields is 
a main ingredient of our model. We now present the simulations based
on these ideas. 
 
\section {Numerical simulations}

We simulate the light curve of the six mini-outbursts and their sudden 
cessation in EG Cnc by means of a simplified model used for the
simulation of the light curves of SU UMa stars 
(Osaki 1989) and WZ Sge (Osaki 1995a). Our model is 
based on a formulation by Anderson (1988) to treat time evolution of 
the disk radius in a dwarf nova outburst cycle. It was used 
extensively by one of the authors to simulate various light curves of SU 
UMa-type dwarf novae (Osaki 1989 for SU UMa stars in general,
Osaki 1995a for WZ Sge, Osaki 1995b for ER UMa,
Osaki 1995c for RZ LMi).  
The model disk consists of an inviscid disk component and a torus 
component.  Since the disk component in our simplified model is taken as
inviscid, only an outside-in type outburst can be treated. The 
outside-in outburst is triggered when the torus mass exceeds its 
critical value. We modify our previous calculations for the WZ Sge
(Osaki 1995a) now taking into account a viscosity decay as described
in the previous section. 

  The binary parameters used for EG Cnc in our simulations are those 
used for WZ Sge in the previous work (Osaki 1995a): (1) the 
mass of the primary white dwarf, $M_1=1.0 M_{\odot}$; (2) the radius of 
the white dwarf, $R_1=0.6\times 10^9$cm;
(3) the mass of the secondary star, $M_2=0.1 M_{\odot}$; (4) the binary 
separation, $A=4.46\cdot 10^{10}$cm; (5) the circularization radius 
or Lubow-Shu radius, $r_{\rm LS}=0.23A=10^{10}$cm. We take a
mass transfer rate $\dot M=1.5 \cdot 10^{15}$gs$^{-1}$.

In order to explain the long quiescence of EG Cnc, 19 years, we 
assume that the quiescent disk viscosity is extremely low, 
$\alpha_{\rm cold} \sim 0.001$. In our simplified model, the viscosity 
appears through the quantity $\beta$,  
$$\beta=(\Delta r/r_t)(\Sigma_{\rm max}/\Sigma_{\rm cool})$$
where $r_t$  is the torus radius, i.e., the radius of the
disk's outer edge and $\Delta r$ the radial width of the torus.
$\Sigma_{\rm max}$ is the critical maximum surface 
density and  $\Sigma_{\rm cool}$ the surface density of the cold disk just 
after the end of the outburst, estimated to be $\Sigma_{\rm 
cool} \simeq 2\times \Sigma_{\rm min}$ (for more 
details see  Osaki 1989).  Therefore, a low viscosity value, $\alpha$, 
corresponds to  a large $\Sigma_{\rm max}$ and therefore to a large
$\beta$. It may be noted here that we used $\beta=0.6$ to simulate
VW Hyi (Osaki 1989) while $\beta=40$ was used to simulate the long
quiescence of WZ Sge (Osaki 1995a). Similarly to explain the long
quiescence of EG Cnc we adopted $\beta=20$ for the pre-outburst cold
disk. The light curve of the main outburst of EG Cnc is calculated in
a similar way as that for WZ Sge  (Osaki 1995a).  

For the post-outburst rebrightening of EG Cnc, we now incorporate the
viscosity decay described in the previous section into our 
simplified model. In order to accommodate the exponential decay of viscosity, 
we take the quantity $\beta$ not constant in time, but exponentially
increasing in the following way 
$$\beta=\beta_0 \exp(t/\tau)$$
where $\beta_0$ is an initial value of $\beta$, and $\tau$ is a 
characteristic time for the viscosity decay, of the order of a week
to a month, and $t$ is the time counted from the end of either the
main outburst or each mini-outburst. That is, viscosity is assumed
to be refreshed to a high value each time a mini-outburst occurs.
In this paper, the quantities $\beta_0$ and $\tau$ are two adjustable
parameters to reproduce the sequence of rapid mini-outbursts with
a mean recurrence time of seven days. This indicates a fairly high viscosity
and requires $\beta_0$ to be small.

\begin{figure}[ht]
\includegraphics[width=8.8cm]{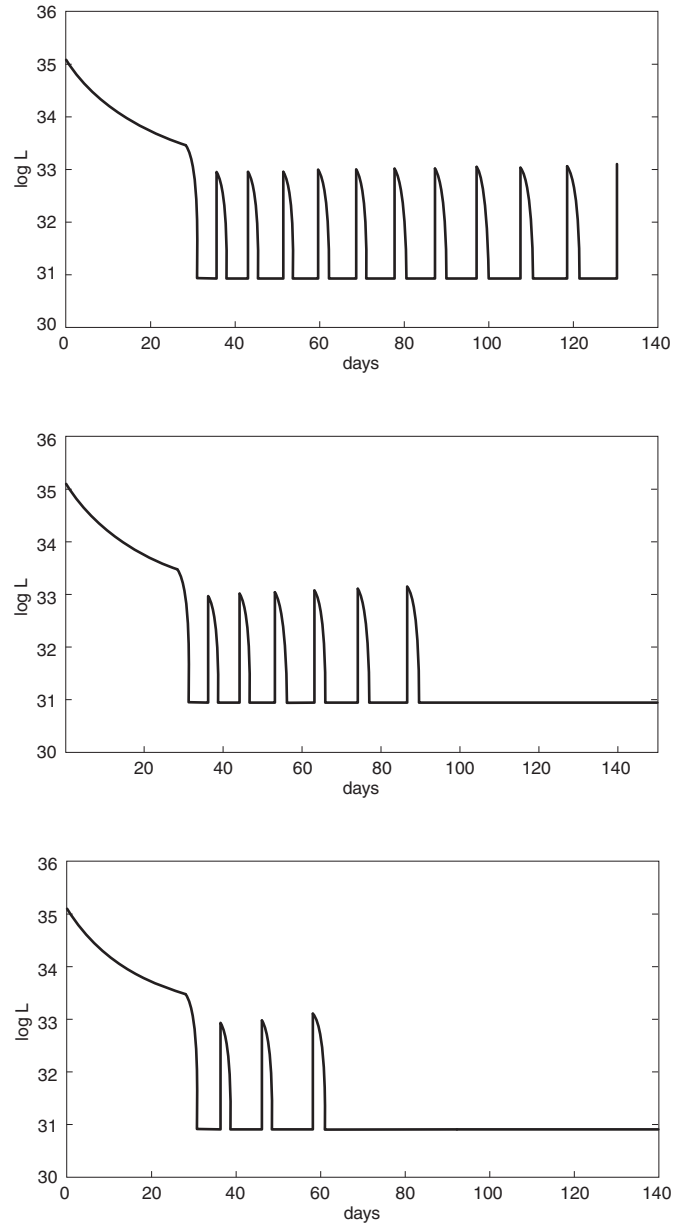}
\caption {Light curve simulations based on the simplified model, for a 
case with $1\%$ increase in $\beta_0$ after each mini-outburst (top 
panel), for $2\%$ increase (middle panel) and for $4\%$ increase 
(bottom panel).}
\end{figure}

We have varied the parameter values for these two quantities and 
found that $\beta_0=0.08$ and $\tau=25$ days reproduce the observed 
recurrence time of 7 days (other choices may also be possible). 
If with each mini-outburst the viscosity in the cold disk is refreshed 
to the original high value the mini-outbursts continue to repeat.
This can be seen from the full simulations with all parameters kept the 
same and with enhanced viscosity in the cold disk (Osaki et al. 1997).
However as discussed in the previous section, 
there will be fluctuations in strength and configuration of 
magnetic fields which are regenerated during each mini-outburst. 
In order to account for such stochastic fluctuations in magnetic
fields and therefore in the resultant viscosity, we have performed
simulations with secular variations of $\beta_0$ and $\tau$.
For instance, $\beta_0$ is not returned back to the same value
but a small increase is allowed after each mini-outburst.
Fig. 4 illustrates three light curves calculated for an
increase of the value $\beta_0$ by $1\%$, $2\%$, and $4\%$ relative to
the previous value when a new mini-outburst occurred. As seen in Fig. 4, 
top panel, more than ten mini-outbursts occur in sequence. With
$2\%$ increase of $\beta_0$ an abrupt cessation after the sixth 
outburst occurs (middle panel). The most important feature of these
calculations is that a different cessation of outbursts occurs for an even 
extremely small difference of the parameters. The light curve shown in the 
middle panel of Fig. 4 with its six consecutive mini-outbursts and their 
sudden end is very similar to that observed in EG Cnc. This sensitive
dependence of the number of mini-outbursts on the parameter $\beta_0$
suggests that the exact number of mini-outbursts is a matter of chance
and can be different after another superoutburst.
Note that the bolometric luminosity shown 
in Fig. 4 is due to radiation from the hot disk plus that from the 
stream impact, and no contributions  are included either from the 
secondary star, or the central white dwarf, or the cold disk.   

This phenomenon of sudden cessation is most easily understood from
Fig. 5 where the evolutions of the torus mass and of its critical
value are shown in the same diagram for the case of $2\%$ increase.
Each time the torus mass reaches its critical value a thermal
instability is triggered and the disk jumps to a hot state. Six times
the increase of torus mass was fast enough to reach the critical
torus mass and outbursts were ignited. However in the last case 
these two curves came close to each other but nearly missed contact
and diverged afterwards. This explains a sudden cessation of outbursts
in our calculations.     

\begin{figure*}[ht]
\includegraphics[width=11.cm]{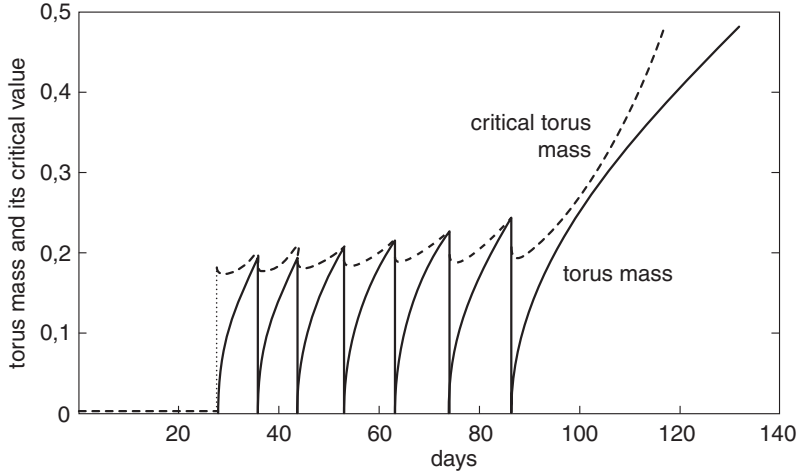}
\caption{Evolution of the torus mass after the superoutburst,
together with that of the critical mass in the torus for a $2\%$
increase in $\beta_0$ (correspondingly decreasing viscosity) after
each mini-outburst, same calculation as shown in the middle panel 
of Fig. 4. } 
\end{figure*}

Similar calculations have been performed by slightly reducing the 
characteristic time $\tau$ after each mini-outburst and we have obtained 
similar results to those shown above. We expect to also get similar results 
if we decrease the mass supply rate slightly after each mini-outburst 
because then it becomes more and more difficult to reach the critical
surface density (see discussion Sect. 4.2).

\section {Discussion}
\subsection {The quiescent luminosity of EG Cnc}
Further support for viscosity decay comes from the decrease of
the quiescent luminosity from 16 to 18 mag within one or two
months after the cessation of the mini-outbursts of EG Cnc (Patterson
et al. 1998, compare Fig. 1). This decrease in quiescent luminosity
is most naturally interpreted as caused by a decrease of luminosity of
the cold disk. We 
first note that most light in minimum of the mini-outbursts comes from 
the cold disk and not from the central white dwarf because in 
quiescence during mini-outbursts the color is very red
(Patterson et al. 1998). A decrease in luminosity after the end of
the rapid mini-outbursts thus indicates a decrease in luminosity of
the cold disk, in accordance with our model.   

Since the radiative flux, $F$, from the surface of an accretion disk in 
thermal equilibrium is given in terms of viscous stresses in the 
standard accretion disk by 
$$2F={9\over 4}\nu \Sigma \Omega={3\over 2}\alpha_{\rm cold} \Omega
\int{Pdz} , $$
we can write for the quiescent disk luminosity, $L_{\rm disk}$,

$$L_{\rm disk}=\int (2F) dS=\int \int {3\over 2} \alpha_{\rm 
cold} (\Re/\mu)T $$

$$ \Omega \rho dz dS  \simeq \alpha_{\rm cold}
{3\over 2}(\Re/\mu)\langle T\rangle \langle \Omega \rangle M_{\rm disk}$$

where $dS$ is the surface element, $P$, $\nu$, $\rho$, $T$, $\Re$, 
$\mu$, $M_{\rm disk}$ are pressure, kinematic viscosity, density, 
temperature, gas constant, mean molecular weight of the gas in the cold 
disk, and disk mass, respectively. Quantities within brackets are 
mean values over the disk. In our light curve simulations 
presented in the previous section, no contribution from the cold disk is 
taken into account, thus we have to add its contribution to the
luminosity of the star in quiescence. Since almost no accretion occurs 
from the disk to the central star during quiescence, the mass in the
disk taken in the equation above should not decrease. The decrease
of luminosity from 16 mag to 18 mag after the end of the mini-outbursts
therefore may result from a decrease of the quantity
$\alpha_{\rm cold}\ \langle T \rangle$. Since the disk 
temperature goes down with a decrease of viscosity, the viscosity 
parameter, $\alpha_{\rm cold}$, must have decreased during this
interval pointing to even lower conductivity.

\subsection {A special feature of WZ Sge stars}
Finally, let us address the question why the post-outburst 
rebrightening phenomenon occurs most often in WZ Sge stars.
This question was addressed before by Kato et al. (1997)
and the following possibility and the observational evidence
supporting it were already put forward by these authors. We
essentially follow their line. One of the distinguishing characters
of WZ Sge stars as compared with other SU UMa stars is the large
amount of mass accumulated in the disk during the long
quiescence. This leads to their outstanding large-amplitude
and long-duration outbursts. This will also bring substantially more
mass beyond the 3:1 resonance radius in order to transfer the large
amount of angular momentum released during accretion. If a fraction of
this is left over when the cooling transition begins to propagate and the
main outburst ends, its viscous relaxation during the beginning
quiescence will feed mass from the outer disk inward at an initially
higher but gradually decreasing rate. This could well be instrumental
to bring about the initial reflaring in the way discussed in this
article, making it peculiar to WZ Sge stars. Evidence for this may be
seen in the observation of superhumps in the early quiescence in EG
Cnc (Kato et al. 1997, Patterson et al. 1998)
pointing to a reservior of mass beyond the 3:1 resonance radius, 
and in the decreasing level of the disk light during the
quiescence in between the sequence of mini-outbursts (Fig. 1), indicating the
gradual depletion of this reservoir of additional mass supply.

\section {Conclusions}
Already in earlier modeling of the light curve of EG Cnc (Osaki et al.
1997) the six mini-outbursts were reproduced in good agreement
with the observations. In these calculations the viscosity in the cold 
disk was kept artificially high during the consecutive mini-outbursts, 
but was assumed to drop to an extremely small value after the end 
of the sixth mini-outburst. These viscosity values were taken ad hoc to fit
the observations, without a specific model for such changes. 
The introduction of magnetic field decay now provides a reason for the
gradual change of viscosity from the high value in the hot state to the
lower and finally very low value in quiescence. These considerations
constitute a physical justification for the assumptions made by
Osaki et al. (1997). Those more detailed calculations can
otherwise be seen as a confirmation of the now performed work based on a
simplified model.  

In the present investigation we have proposed a model for
the consecutive mini-outbursts of EG Cnc and their abrupt cessation
based on the decay of the small-scale-dynamo produced MHD turbulent
viscosity due to finite conductivity of the cold disk matter.
We have successfully simulated the light curve of EG Cnc using a
simplified model that can treat only the outside-in outburst,
that is, outbursts occur when the torus reaches its critical value.
In reality the ignition of mini-outbursts in EG Cn 
might occur most likely in the middle part and not
necessarily near the outer edge of the disk. In such a case, what was
summarily called `mass supply from the secondary star' in our
simulations may be
interpreted as mass supply by viscous diffusion from the outer disk
to the ignition point. 

Should this model presented here prove correct it lends strong support
to the theory of magnetohydrodynamic turbulence as cause of the
viscosity in high temperature accretion disks. It also supports again
the notion of very low viscosity for the quiescent disks in WZ Sge
stars as compared to that of ordinary dwarf novae. In this and in other
features not discussed here in detail it fits well into the gradually
emerging solution of the puzzle of accretion disk viscosity.

\begin{acknowledgements}
Yoji Osaki acknowledges financial support from
the Japanese Ministery of Education, Science, Sports and Culture with
a Grants-in Aid for Scientific Research No.12640237.
\end{acknowledgements}

\end{document}